\newif\iftwocolumn\twocolumntrue

\documentclass{IEEEoj}
\usepackage{cite}
\usepackage{amsmath,amssymb,amsfonts}
\usepackage{algorithmic}
\usepackage{graphicx,color}
\usepackage{textcomp}

\ifCLASSINFOpdf
\else
\fi

\usepackage{amsmath,stackengine}
\usepackage{amssymb}
\usepackage{bm}
\usepackage{amsthm}
\usepackage{dsfont}
\usepackage{siunitx}

\newtheorem{remark}{Remark}
\newtheorem{lemma}{Lemma}
\usepackage{xcolor}

\usepackage{subfig}

\DeclareRobustCommand*{\IEEEauthorrefmark}[1]{%
  \raisebox{0pt}[0pt][0pt]{\textsuperscript{\footnotesize #1}}%
}

\makeatletter
\newcommand{\customlabel}[2]{%
\protected@write \@auxout {}{\string \newlabel {#1}{{#2}{}}}}
\makeatother

\def\OJlogo{}
\begin{document}

\title{Approximation-based Threshold Optimization from Single Antenna to Massive SIMO Authentication}

\author{\IEEEauthorblockN{Stefan Roth\IEEEauthorrefmark{1}, Aydin Sezgin\IEEEauthorrefmark{1}, Roman Bessel\IEEEauthorrefmark{1} and H. Vincent Poor\IEEEauthorrefmark{2}}}
\affil{Institute of Digital Communication Systems, Ruhr University Bochum, Bochum, Germany}
\affil{Department of Electrical and Computer Engineering, Princeton University, Princeton, NJ}
\corresp{CORRESPONDING AUTHOR: Stefan Roth (e-mail: stefan.roth-k21@rub.de).}
\authornote{This work has been funded by the Deutsche Forschungsgemeinschaft (DFG, German Research Foundation) under Germany's Excellence Strategy - EXC 2092 CASA - 390781972.}
\markboth{Kalman-based PHY Authentication: System Analysis and Threshold Optimization}{Roth \textit{et al.}}

\begin{abstract}
In a wireless sensor network, data from various sensors are gathered to estimate the system-state of the process system. However, adversaries aim at distorting the system-state estimate, for which they may infiltrate sensors or position additional devices in the environment. To authenticate the received process values, the integrity of the measurements from different sensors can be evaluated jointly with the temporal integrity of channel measurements from each sensor. For this purpose, we design a security protocol, in which Kalman filters are used to predict the system-state and the channel-state values, and the received data are authenticated by a hypothesis test. We theoretically analyze the adversarial success probability and the reliability rate obtained in the hypothesis test in two ways, based on a chi-square approximation and on a Gaussian approximation. The two approximations are exact for small and large data vectors, respectively. The Gaussian approximation is suitable for analyzing massive single-input multiple-output (SIMO) setups. To obtain additional insights, the approximation is further adapted for the case of channel hardening, which occurs in massive SIMO fading channels. As adversaries always look for the weakest point of a system, a time-constant security level is required. To provide such a service, the approximations are used to propose time-varying threshold values for the hypothesis test, which approximately attain a constant security level. Numerical results show that a constant security level can only be achieved by a time-varying threshold choice, while a constant threshold value leads to a time-varying security level.
\end{abstract}


\maketitle

\IEEEpeerreviewmaketitle

\section{Introduction}
\IEEEPARstart{I}{n} wireless sensor networks and networked control systems, various process systems need to be monitored at remote devices. Therefore, the sensor information is transmitted over the network and gathered at a remote device, where these are utilized for process monitoring and control purposes. 
Thereby, the validity of the received data can be crucial for the functionality of such applications. It is thus critical to design the communication in wireless sensor networks reliably and securely. Security-disrupting attacks from adversaries can come in various forms, such as replay attacks, false-data injection (FDI) attacks, and denial-of-service (DoS) attacks \cite{7011201}. In the case of FDI attacks, they can either pertain to the physical manipulation of a sensor at the device or spoofing at the receiver. As a countermeasure, the receiver needs to authenticate the received data packets.

Therefore, authentication schemes have been proposed, which rely on three different phases: A registration phase, a login phase, and an authentication phase \cite{1636182,4411100}. The registration phase is used to introduce the devices to each other by initially exchanging secrets between the devices. In the login phase, the secrets are used further to exchange information, which is later used to authenticate a series of transmissions.
To protect a system against attackers, it is essential to secure the weakest point of a system \cite{DBLP:books/daglib/0023872}. This means that there is a lower bound of the security level, which the authentication system needs to ensure constantly over time. Note that threshold-based authentication schemes show a trade-off between the achieved security and reliability. Hence, here we specifically focus on designing the authentication process such that the achieved security level is constant over time. To fulfill this condition, the threshold value needs to be optimized for each received packet individually.

\begin{figure}
    \centering
    \includegraphics{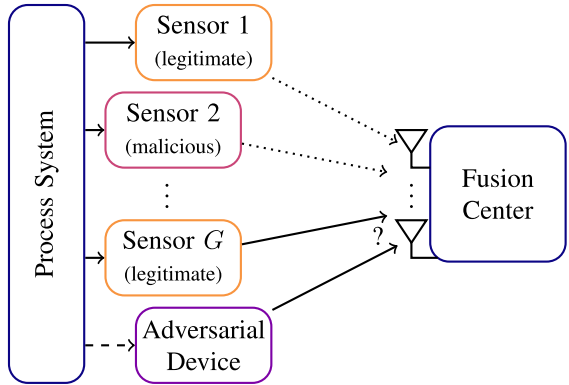}
    \caption{The fusion center wants to authenticate legitimate sensor data in the presence of malicious sensors and adversarial devices.}
    \label{fig:motivation}
\end{figure}

This work focuses on authenticating data and assuring system integrity by detecting FDI attacks from adversarial devices or malicious sensor nodes at the physical layer. Therefore, the authenticity is validated by employing measurements of the channel and the sensors jointly. We consider the communication between multiple sensors and a fusion center as shown in \figurename~\ref{fig:motivation}, in which the data are authenticated by a combination of these two measures.
\begin{itemize}
    \item Multiple sensors measure data from the same physical process or control system. By checking the integrity of the measurements from the different sensors, malicious sensor nodes can be detected.
    \item The channel values are measured regularly within existing physical layer communication protocols. These values are highly location-dependent and cannot fully be replicated by a spatially distinct adversary. By evaluating the temporal integrity of the channel measurements, adversarial transmitters added to the surroundings can be detected.
\end{itemize}
\begin{figure*}
\centering
\includegraphics{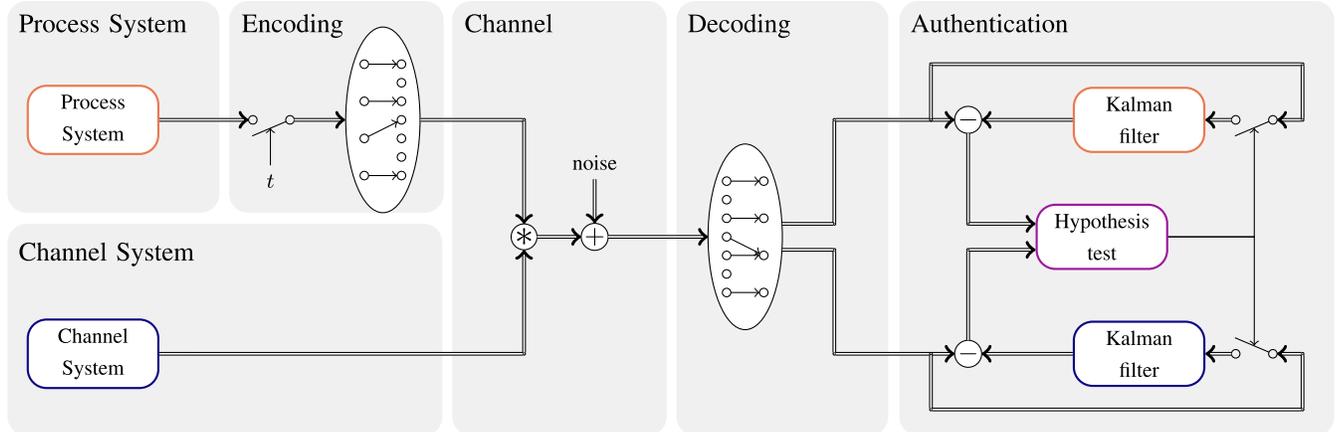}
\caption{The received data are authenticated based on two measures: The values of the process system and the channel system.}
\label{fig:schematic}
\end{figure*}

Between different internet of things (IoT) devices and wireless sensor networks, the number of available resources varies. Some sensors and fusion centers are low-cost and equipped with only a single antenna. Other fusion centers are connected to a base station, which may be equipped with a massive number of antennas and thus able to apply massive single-input multiple-output (SIMO) techniques \cite{BANA2019100859}. Also, the number of measurements obtained from each sensor might vary. While some sensors only measure a single value (such as temperature or filling level sensors), other sensors measure various types of information together (such as image sensors). This means that authentication schemes are needed that are suitable to secure systems with a wide range of available resources.
When the number of resources is large, channel hardening can become relevant \cite{1327795}. While the channel and sensor measurement of a single antenna or sensor can have high statistical variations, the effect of the variations is significantly reduced when a large number of resources is considered. This effect has been investigated for the channel gain obtained in massive SIMO systems with fading channels in \cite{7880691}. For threshold-based authentication systems, channel hardening will lead to a significant reduction of the time-dependency of the different optimized threshold values.

In the following, we will provide an overview of related authentication methods.

\subsection{Literature Review}

As protection against FDI and replay attacks, different kinds of physical layer authentication schemes have been considered in the literature. Thereby, various kinds of approaches have been employed to provide authenticity and attack-resilience. Especially the stochastic nature of wireless channels, fingerprints of transmitter hardware impairments, and watermarking methods have been widely used for authentication \cite{8920228}. Additionally, there has been a widespread research focus on attack-resilient state estimation \cite{7563303} and assuring data integrity \cite{7403134}. For all of these kinds of approaches, similar techniques have been used to develop authentication schemes.

As examples of \emph{hardware impairments}, the carrier frequency offset and the in-phase/quadrature imbalance (IQI) have been employed for authentication in \cite{6804410} and \cite{7036875}, respectively. Thereby, in \cite{6804410} the carrier frequency offset has been followed by a Kalman filter to provide authenticity. In \cite{7036875}, IQI has been used to design an authentication scheme for an amplify-and-forward setup.

The properties of the \emph{channel} have been exploited to authenticate devices. Therefore, the channel measurements have been classified based on various signal processing and machine learning techniques, such as support vector machines (SVMs) \cite{7924970}, Kalman filters \cite{9034987,9413408}, and generative adversarial neural networks (GANs) \cite{9310070}. A channel-based authentication scheme for massive multiple-input multiple-output (MIMO) systems has been proposed in \cite{8920228}.
Moreover, the channel measurements can be processed to obtain information on the transmitter location and the surroundings. In \cite{8640074}, the angle of arrival has been used to authenticate the transmitter in a vehicular communication network (V2X) scenario.

Moreover, various other works have focused on securing control systems based on validating the \emph{integrity of measurements} from monitored or controlled systems. For example, the integrity of data from multiple sensors is evaluated to find malicious sensors in \cite{8758160}. In \cite{7470566}, a chi-square detector is established to tackle the issue of FDI attacks in the form of linear manipulation of the sensor values in the case of a single sensor.

Especially \emph{Kalman-filter}-based algorithms have been widely utilized in the context of integrity validation. Here, cases range from authentication based on transmitter hardware \cite{6804410} and the channel \cite{9034987,9413408} to validating the integrity of the measurements from different sensor nodes \cite{8758160,7470566}. Nevertheless, there has only been very limited research on using channel and process parameters \emph{jointly} for authentication.

\subsection{Contributions}
In this work, we apply a Kalman-filter-based scheme to authenticate the data transmission in a wireless sensor network. The contributions are as follows:
\begin{itemize}
    \item While other works have focused on authenticating the users either based on the channel or the process parameters, we employ both kinds of parameters \emph{jointly} to create a secure and attack-resilient state-estimation system.
    \item We provide three \emph{approximations for the security and the reliability level} obtained under a given threshold. The first approximation holds exactly for a small amount of resources, such as very few antennas and sensors. The second approximation instead is exact for an asymptotically large amount of resources as available in massive SIMO. The third approximation takes the effect of channel hardening into account and is otherwise similar to the second.
    \item The approximations are then used to suggest \emph{time-dependent thresholds} that enable the approximate fulfillment of a constant security level. For a single resource and an asymptotically large number of resources, the security level can be fulfilled exactly in the considered model.
\end{itemize}

\subsection{Notation} Vectors $\bm{x}$ and matrices $\bm{X}$ are denoted in bold lower-case and bold upper-case letters, respectively. $(\cdot)^*$ and $(\cdot)^H$ indicate the conjugate and conjugate-transpose, respectively.  $\mathcal{N}(\bm{\underline{x}},\bm{X})$ and $\mathcal{CN}(\bm{\underline{x}},\bm{X})$ refer to the real-valued and complex-valued Gaussian density function, respectively, with mean $\bm{\underline{x}}$ and covariance $\bm{X}$; and $Q(x)=\int_{x}^{\infty}\frac{1}{\sqrt{2\pi}}\exp\left(-\frac{y^2}{2}\right)dy$ refers to the $Q$-function. Moreover, $Q_M\left(a,b\right)$ and $Q(M,b)$ refer to the Marcum $Q$-function and the upper regularized incomplete gamma function, respectively.

\section{System Model}
The considered system consists of $G$ sensors, which observe the process system and transmit measurements via the wireless medium to the fusion center regularly. In \figurename~\ref{fig:schematic}, the interaction of one sensor and the fusion center is shown schematically.
\begin{itemize}
    \item Each sensor regularly measures data from the process system and transmits these to the fusion center.
    \item For the transmission, a channel coding scheme is used, which allows for jointly decoding the sensor observation and measuring the channel.
    \item The fusion center uses the two observations to authenticate the received data.
\end{itemize}
Next, we focus on the model of the process system and the channels. Afterward, we will discuss the adversary model.
\subsection{Process System}
We focus on a process system whose system-state values $\bm{x}_{\mathrm{P}}(t)$ are Gauss-Markov distributed \cite{10.5555/1816978}. This means that the system-state values are modeled by a state-space equation and slowly change over time. Thereby, the derivative of the system-state values depends linearly on the system-state values. Hence, the model of the process system can be described as
\begin{subequations}\label{eq:x_p_sysmodel}
\begin{align}
    \frac{\partial\bm{x}_{\mathrm{P}}(t)}{\partial t}&=\bm{A}_{\mathrm{P}}\bm{x}_{\mathrm{P}}(t)+\bm{u}_{\mathrm{P}}(t),\\
    \bm{x}_{\mathrm{P}}(0)&=\bm{x}_{\mathrm{P},0}.
\end{align}
\end{subequations}
Thereby, the input noise $\bm{u}_{\mathrm{P}}(t)$ is distributed as $\bm{u}_{\mathrm{P}}(t)\sim\mathcal{CN}(\bm{0},\bm{U}_{\mathrm{P}})$. 
We are considering the case that the state matrix is diagonalizable with an eigenvalue decomposition of $\bm{A}_{\mathrm{P}}=\bm{W}_{\mathrm{P}}\bm{\Lambda}_{\mathrm{P}}\bm{W}_{\mathrm{P}}^{-1}$ and the system is stable, i.e., $\lambda_{\mathrm{P},i}=\left(\bm{\Lambda}_{\mathrm{P}}\right)_{i,i}<0$. In this case, the system is stationary and the covariance matrix of the system-state $\bm{x}_{\mathrm{P}}(t)$ at time $t$ can be derived as \cite{9646490}
\begin{align}
    \bm{X}_{\mathrm{P}}=\bm{W}_{\mathrm{P}}\bm{\ddot{X}}_{\mathrm{P}}\bm{W}_{\mathrm{P}}^{H},
\end{align}
where
\begin{align}
    \left(\bm{\ddot{X}}_{\mathrm{P}}\right)_{i,j}&=-\frac{\left(\bm{W}_{\mathrm{P}}^{-1}\bm{U}_{\mathrm{P}}\bm{W}_{\mathrm{P}}^{-H}\right)_{i,j}}{\lambda_{\mathrm{P},i}+\lambda_{\mathrm{P},j}^*}.
\end{align}

Each sensor $g$ observes only the part of the system described by the output matrix  $\bm{C}_{\mathrm{P},g}$ and transmits the measurement at time instants $t$ via the channel to the fusion center. Due to channel noise and channel coding, an output noise $\bm{n}_{\mathrm{P},g}(t)\sim\mathcal{CN}(\bm{0},\bm{N}_{\mathrm{P}})$ is induced. Hence, the sensor measurement received by the fusion center is the $M_{\mathrm{P}}$-dimensional vector specified by the output equation
\begin{align}
    \bm{y}_{\mathrm{P},g}(t)=\bm{C}_{\mathrm{P},g}\bm{x}_{\mathrm{P}}(t)+\bm{n}_{\mathrm{P},g}(t).
\end{align}

\subsection{Channel Model}

The sensor nodes alternately record new measurements and transmit them via the wireless medium to a fusion center. All sensor nodes are equipped with a single antenna, while the fusion center is connected to a base station with $M_{\mathrm{C}}$ antennas.
We consider fading channels, in which the channel-states $\bm{x}_{\mathrm{C},g}(t)$ follow a Gauss-Markov model, i.e., the state values develop as
\begin{subequations}\label{eq:x_c_sysmodel}
\begin{align}
    \frac{\partial\bm{x}_{\mathrm{C},g}(t)}{\partial t}&=\bm{A}_{\mathrm{C}}\bm{x}_{\mathrm{C},g}(t)+\bm{u}_{\mathrm{C},g}(t),\\
    \bm{x}_{\mathrm{C},g}(0)&=\bm{x}_{\mathrm{C},g,0}.
\end{align}
\end{subequations}
Here, $\bm{A}_{\mathrm{C}}$ is the state matrix of the channel, while the input noise $\bm{u}_{\mathrm{C},g}(t)$ is distributed as $\bm{u}_{\mathrm{C},g}(t)\sim\mathcal{CN}(\bm{0},\bm{U}_{\mathrm{C}})$. Hence, the channel measurement of legitimate packets is
\begin{align}
    \bm{y}_{\mathrm{C},g}(t)&=\bm{C}_{\mathrm{C},g}\bm{x}_{\mathrm{C},g}(t)+\bm{n}_{\mathrm{C},g}(t),
\end{align}
of the size $M_{\mathrm{C}}$, in which $\bm{C}_{\mathrm{C},g}$ is the output matrix of the channel, while the output noise is $\bm{n}_{\mathrm{C},g}(t)\sim\mathcal{CN}(\bm{0},\bm{N}_{\mathrm{C}})$.
Due to the proximity of the antennas, the entries of the output vector $\bm{y}_{\mathrm{C},g}$ can be correlated. To reflect this, the matrices $\bm{U}_{\mathrm{C}}$, $\bm{A}_{\mathrm{C}}$ and $\bm{C}_{\mathrm{C},g}$ can have an arbitrary structure.
We assume that the state and output matrix, as well as the respective noise covariance matrix for both the system and the channel, are known to all parties. Moreover, the channel coding is designed such that the fusion center can measure the channel output $\bm{y}_{\mathrm{C},g}(t)$ from the received signal in addition to the transmitted process measurement $\bm{y}_{\mathrm{P},g}(t)$.

\subsection{Adversary Model}
Previously, we have modeled the transmission of legitimate transmitters as shown in \figurename~\ref{fig:schematic}. However, there might also be adversaries present, whose motive is to get their packets authenticated by the receiving actuator. To model these, we are assuming two kinds of adversaries as shown in \figurename~\ref{fig:motivation}:
\begin{itemize}
\item Adversaries can position their own devices in the environment. Due to their distinct locations, they typically have individual channels, which are different from those of legitimate sensors \cite{7849064}. Hence, a strategy to mitigate further deviations might be to first try to mimic the system-state transmitted from the legitimate sensor, before transmitting adversarial data.
\item Also, attackers might infiltrate existing sensors. In this case, the sensor will transmit a malicious system observation, while the measured channel-state will match the one of the previous transmission. 
\end{itemize}
In the general case, a packet transmitted by the adversary will contain a partially different system-state value, which is modelled to be distributed as
\begin{align}
    \bm{\tilde{y}}_{\mathrm{P},g}(t)\sim\mathcal{CN}\left(\bm{\tilde{\underline{y}}}_{\mathrm{P},g}(t),\bm{\tilde{Y}}_{\mathrm{P},g}(t)\right).
\end{align}
Thereby, $\bm{\tilde{\underline{y}}}_{\mathrm{P},g}(t)$ and $\bm{\tilde{Y}}_{\mathrm{P},g}(t)$ are the mean and covariance of the received data, which can be correlated to the vector $\bm{y}_{\mathrm{P},g}(t)$ from the legitimate sensor.

Moreover, we assume the receiver to measure a channel from the adversary, which is distributed as the Gaussian
\begin{align}
    \bm{\tilde{y}}_{\mathrm{C},g}(t)\sim\mathcal{CN}\left(\bm{\tilde{\underline{y}}}_{\mathrm{C},g}(t),\bm{\tilde{Y}}_{\mathrm{C},g}(t)\right),
\end{align}
which has a mean and covariance matrix of $\bm{\tilde{\underline{y}}}_{\mathrm{C},g}(t)$ and $\bm{\tilde{Y}}_{\mathrm{C},g}(t)$. Thereby, the measurement noise occuring at the receiver is already included in $\bm{\tilde{Y}}_{\mathrm{P},g}(t)$ and $\bm{\tilde{Y}}_{\mathrm{C},g}(t)$.

This means that depending on whether the packet has been transmitted by the legitimate or adversarial transmitter, the received process and channel measurements are
\begin{align}
    \bm{y}_{\mathrm{P},g}(t)&=\begin{cases}\bm{C}_{\mathrm{P},g}\bm{x}_{\mathrm{P}}(t)+\bm{n}_{\mathrm{P},g}(t)&\text{legitimate transmitter}\\\bm{\tilde{y}}_{\mathrm{P},g}(t)&\text{adversarial transmitter}\end{cases},\label{eq:y_p_options}\\
    \bm{y}_{\mathrm{C},g}(t)&=\begin{cases}\bm{C}_{\mathrm{C},g}\bm{x}_{\mathrm{C},g}(t)+\bm{n}_{\mathrm{C},g}(t)\hspace{-0.5cm}\ \\&\text{legitimate transmitter}\\\bm{\tilde{y}}_{\mathrm{C},g}(t)&\text{adversarial transmitter}\end{cases}.\label{eq:y_c_options}
\end{align}
Thereby, adversarial devices positioned in the surrounding might choose $\bm{\tilde{y}}_{\mathrm{P},g}(t)=\bm{C}_{\mathrm{P},g}\bm{x}_{\mathrm{P}}(t)+\bm{n}_{\mathrm{P},g}(t)$ to optimize the probability of getting a packet accepted, while the channel measurement is distributed as $\bm{\tilde{y}}_{\mathrm{C},g}(t)\sim\mathcal{CN}\left(\bm{0},\bm{C}_{\mathrm{C},g}\bm{X}_{\mathrm{C}}\bm{C}_{\mathrm{C},g}^H+\bm{N}_{\mathrm{C}}\right)$. However, an infiltrated sensor typically has a channel equaling the one of a legitimate transmission, i.e., $\bm{\tilde{y}}_{\mathrm{C},g}(t)=\bm{C}_{\mathrm{C},g}\bm{x}_{\mathrm{C},g}(t)+\bm{n}_{\mathrm{C},g}(t)$, which is highly correlated to the channel from the previous transmission. In this case, the attacker might choose a sensor value of $\bm{\tilde{y}}_{\mathrm{P},g}(t)\sim\mathcal{CN}\left(\bm{0},\bm{C}_{\mathrm{P},g}\bm{X}_{\mathrm{P}}\bm{C}_{\mathrm{P},g}^H+\bm{N}_{\mathrm{P},g}\right)$. In the following, we will design a physical layer security protocol, which secures the system against both kinds of attacks.

\section{Protocol Design}
Due to the potential presence of adversaries, it is essential to authenticate the received data. Therefore, we design a security protocol separated into a registration phase, a login phase, and an authentication phase, in which we specifically focus on the latter.

Before the arrival of the first packet at time $t$, the system-state and channel-state values can be predicted as
\begin{align}
    \bm{\hat{\underline{z}}}_{\mathrm{P}}(t|-1)&=\bm{0},&\bm{\hat{\underline{z}}}_{\mathrm{C},g}(t|-1)&=\bm{0}.\label{eq:z_mean_init}
\end{align}
Thereby, the uncertainties of the predictions are
\begin{align}
    \bm{\hat{Z}}_{\mathrm{P}}(t|-1)&=\bm{X}_{\mathrm{P}},&\bm{\hat{Z}}_{\mathrm{C},g}(t|-1)&=\bm{X}_{\mathrm{C}}.\label{eq:z_var_init}
\end{align}
In the following, we elaborate on the three phases of the security protocol.

\subsection{Registration Phase}

When a new sensor logs in to the system, the first packet needs to be verified. Therefore, a trusted environment can be set up to secure the corresponding packets. Alternatively, the sensor is registered in advance to the fusion center by establishing a common secret between the devices \cite{7270404}.

\subsection{Login Phase}
In the login phase, the sensor transmits the observation of the process system at time $t$ such that the fusion center can verify the transmission with already existing methods. Therefore, one option is to ensure that the environment is trusted during the transmission of these packets, such as in the case of a manual pairing. Alternatively, a common secret can secure the transmitted data if priorly is established between the devices \cite{DBLP:books/daglib/0023872}. 
Afterward, the data included in the received packet are used to update the estimates of the system-state and the channel-state.

Recall that the process system behind all sensor observations is identical. Hence, we focus on employing the difference between the output vectors $\bm{y}_{\mathrm{P},g}(t)$ and the prediction obtained from the most recently authenticated packet from any sensor. The timestamp of the corresponding observation is referred to as $t_{\mathrm{P}}$, which is set to $-1$ for the first observation. As the channel measurements from the different sensors are independent, each channel measurement $\bm{y}_{\mathrm{C},g}(t)$ needs to be compared to a prediction obtained from the most recently authenticated packet from the same sensor, which has been received at time $t_{\mathrm{C},g}$. In the login phase, no prior measurement from the same sensor is available, which is indicated by $t_{\mathrm{C},g}=-1$.

Having these definitions available, the data from the received packet can be used to update the estimates of system-state and channel-state. Therefore, the estimation error, i.e., the difference between measurement and prediction, is required, which is
\begin{align}
    \bm{e}_{\mathrm{P},g}(t)&=\bm{y}_{\mathrm{P},g}(t)-\bm{C}_{\mathrm{P},g}\bm{\hat{\underline{z}}}_{\mathrm{P}}(t|t_{\mathrm{P}}),\label{eq:e_P}\\
    \bm{e}_{\mathrm{C},g}(t)&=\bm{y}_{\mathrm{C},g}(t)-\bm{C}_{\mathrm{C},g}\bm{\hat{\underline{z}}}_{\mathrm{C},g}(t|t_{\mathrm{C},g}).\label{eq:e_C}
\end{align}
As the system-state and the channel-states follow a similar behavior, we focus on developing a Kalman filter \cite{kalman} to track the state of the process system. The signal processing for the channel-states can be described similarly. 
Employing priorly available estimates such as \eqref{eq:z_mean_init} and \eqref{eq:z_var_init}, the received process measurement can be used to obtain an initial estimate of the system-state value of the process and the corresponding uncertainty. Therefore, Kalman filter equations can be used, i.e.,
\begin{align}
    \bm{K}_{\mathrm{P}}(t)&=\bm{\hat{Z}}_{\mathrm{P}}(t|t_{\mathrm{P}})\bm{C}_{\mathrm{P},g}^H\left(\bm{C}_{\mathrm{P},g}\bm{\hat{Z}}_{\mathrm{P}}(t|t_{\mathrm{P}})\bm{C}_{\mathrm{P},g}^H+\bm{N}_{\mathrm{P}}\right)^{-1}\label{eq:Kalman1}\\
    \bm{\hat{\underline{z}}}_{\mathrm{P}}(t)&=\bm{\hat{\underline{z}}}_{\mathrm{P}}(t|t_{\mathrm{P}})+\bm{K}_{\mathrm{P}}(t)\bm{e}_{\mathrm{P},g}(t)\label{eq:Kalman2}\\
    \bm{\hat{Z}}_{\mathrm{P}}(t)&=\left(\bm{I}-\bm{K}_{\mathrm{P}}(t)\bm{C}_{\mathrm{P},g}\right)\bm{\hat{Z}}_{\mathrm{P}}(t|t_{\mathrm{P}}).\label{eq:Kalman3}
\end{align}
Similar expressions with concatenated matrices and vectors hold if measurements from multiple sensors are processed jointly at the same time. The following packets will be authenticated based on the obtained estimates of the system-state and the channel-state.

\subsection{Authentication Phase}
To verify the following receptions, the available estimates of the system-state and the channel-states are used to predict the received signals. The predictions are then compared to the new measurements to authenticate the received data.

Having an estimate from time $t_{\mathrm{P}}$ present, the Kalman filter can be used to predict the system-state at a later time instant $t$ as
\begin{align}
    \bm{\hat{\underline{z}}}_{\mathrm{P}}(t|t_{\mathrm{P}})&=e^{\bm{A}_{\mathrm{P}}(t-t_{\mathrm{P}})}\bm{\hat{\underline{z}}}_{\mathrm{P}}(t_{\mathrm{P}}),\label{eq:Kalman4}\\
    \bm{\hat{Z}}_{\mathrm{P}}(t|t_{\mathrm{P}})&=e^{\bm{A}_{\mathrm{P}}(t-t_{\mathrm{P}})}\bm{\hat{Z}}_{\mathrm{P}}(t_{\mathrm{P}})e^{\bm{A}_{\mathrm{P}}^H(t-t_{\mathrm{P}})}\nonumber\\&\hspace{2.5cm}+\int_{0}^{t-t_{\mathrm{P}}}e^{\bm{A}_{\mathrm{P}}t'}\bm{U}_{\mathrm{P}}e^{\bm{A}_{\mathrm{P}}^Ht'}dt'\nonumber\\
    &=e^{\bm{A}_{\mathrm{P}}(t-t_{\mathrm{P}})}\bm{\hat{Z}}_{\mathrm{P}}(t_{\mathrm{P}})e^{\bm{A}_{\mathrm{P}}^H(t-t_{\mathrm{P}})}+\bm{X}_{\mathrm{P}}-\bm{X}_{\mathrm{P}}^{(t-t_{\mathrm{P}})},
\end{align}
in which $\bm{X}_{\mathrm{P}}^{(t-t_{\mathrm{P}})}=\bm{W}_{\mathrm{P}}\bm{\ddot{X}}_{\mathrm{P}}^{(t-t_{\mathrm{P}})}\bm{W}_{\mathrm{P}}^H$ and
\begin{align}
    \left(\bm{\ddot{X}}_{\mathrm{P}}^{(t-t_{\mathrm{P}})}\right)_{i,j}&=-e^{(\lambda_{\mathrm{P},i}+\lambda_{\mathrm{P},j}^*)(t-t_{\mathrm{P}})}\frac{\left(\bm{W}_{\mathrm{P}}^{-1}\bm{U}_{\mathrm{P}}\bm{W}_{\mathrm{P}}^{-H}\right)_{i,j}}{\lambda_{\mathrm{P},i}+\lambda_{\mathrm{P},j}^*}.
\end{align}
Recall that $\bm{W}_{\mathrm{P}}$ and $\lambda_{\mathrm{P},i}$ are the eigenvector matrix and the $i$-th eigenvalue of $\bm{A}_{\mathrm{P}}$, respectively. We assume that measurements from multiple sensors can be received simultaneously. The predictions for the system-state $\bm{\hat{\underline{z}}}_{\mathrm{P}}(t|t_{\mathrm{P}})$ and the channel-states $\bm{\hat{\underline{z}}}_{\mathrm{C},g}(t|t_{\mathrm{C},g})$ at time $t$ can be used to authenticate all packets arriving at that time $t$. For each packet, we can formulate the two hypotheses
\begin{align}
\begin{cases}
    \mathcal{H}_0&\text{legitimate transmitter}\\
    \mathcal{H}_1&\text{adversarial transmitter}
\end{cases}.
\end{align}
When employing \eqref{eq:e_P} and \eqref{eq:e_C}, a hypothesis test can be formulated to evaluate the packets as
\begin{align}
    \begin{cases}
        \mathcal{H}_0\hspace{-0.24cm}\ & \bm{e}_{\mathrm{P},g}^H(t)\bm{V}_{\mathrm{P},g}\bm{e}_{\mathrm{P},g}(t) < \eta_{\mathrm{P}} \land \bm{e}_{\mathrm{C},g}^H(t)\bm{V}_{\mathrm{C},g}\bm{e}_{\mathrm{C},g}(t) < \eta_{\mathrm{C}}\\
        \mathcal{H}_1\hspace{-0.24cm}\ & \bm{e}_{\mathrm{P},g}^H(t)\bm{V}_{\mathrm{P},g}\bm{e}_{\mathrm{P},g}(t) > \eta_{\mathrm{P}} \lor \bm{e}_{\mathrm{C},g}^H(t)\bm{V}_{\mathrm{C},g}\bm{e}_{\mathrm{C},g}(t) > \eta_{\mathrm{C}}
    \end{cases}.\label{eq:hyp0}
\end{align}
Thereby, $\bm{V}_{\mathrm{P},g}$ and $\bm{V}_{\mathrm{C},g}$ are positive semidefinite weight matrices and $\eta_{\mathrm{P}}$ and $\eta_{\mathrm{C}}$ are the threshold values. When the authentication of some data packets is successful, the corresponding packets are used to enhance the system-state estimate as described in \eqref{eq:Kalman1}, \eqref{eq:Kalman2}, and \eqref{eq:Kalman3}.
\begin{remark}
If there is only a single sensor, the system-state value will only partially be helpful for authentication. If the only sensor is infiltrated, the attacker can change the transmitted channel estimate slowly to remain undetected. In this case, the channel-state measurements can still be used for authentication in the presence of adversarial devices.
\end{remark}
In the next section, we will analyze the hypothesis test results and elaborate on the choice of the threshold variables.

\section{Theoretical Analysis}
In the following, we will analyze the result of the hypothesis test in \eqref{eq:hyp0} analytically. Therefore, we will approximate the number of true and false negatives based on two different approximations, which hold for measurement vectors of different sizes. For low-dimensional systems, the results are based on approximating the left-hand sides as non-central chi-square distributions, while for high-dimensional systems, the left-hand sides are approximated as Gaussian. Afterward, a third approximation is created, which takes the effects of channel hardening occurring in massive SIMO into account. These results will then be used to engineer the thresholds $\eta_{\mathrm{P}}$ and $\eta_{\mathrm{C}}$, as indicated in Remark~\ref{remark:eta}.
\begin{remark}\label{remark:eta}
The thresholds $\eta_{\mathrm{P}}$ and $\eta_{\mathrm{C}}$ connect the reliability, i.e., the number of true negatives $P_{\mathrm{TN}}(t)$, and the security, i.e., the number of false negatives $P_{\mathrm{FN}}(t)$. Hence, a reliability constraint can be converted directly into a level of security for each data packet - while a security constraint can be converted directly into a level of reliability. Alternatively, the thresholds can be chosen such that a function of both probabilities is optimized for each packet. This also allows for designing systems that intelligently apply additional layers of security or reliability only for certain packets.
\end{remark}

\subsection{Chi-Square Approximation for small $M_{\mathrm{P}}$ and $M_{\mathrm{C}}$}\label{sec:chiSquare}
Some IoT devices and sensors have a low number of antennas and observe only a low number of measurements. In these cases, $M_{\mathrm{P}}$ and $M_{\mathrm{C}}$ are low and the probability to remain below the threshold of a simplified version of \eqref{eq:hyp0} is approximated in the following Lemma~\ref{lem:chisquare}.
\begin{lemma}\label{lem:chisquare}
If $\bm{l}\sim\mathcal{CN}\left(\bm{\underline{l}},\bm{L}\right)$ is an $M$-dimensional vector, the statement $\bm{l}^H\bm{l}<\eta$ holds true with probability
\begin{align}
    P=1-Q_M\left(\sqrt{\frac{2}{\Bar{L}}\bm{\underline{l}}^H\bm{\underline{l}}},\sqrt{\frac{2\eta}{\Bar{L}}}\right),\label{eq:cdf_value}
\end{align}
approximately, where $Q_M\left(a,b\right)$ is the Marcum $Q$-function \cite{1057560}. Thereby, $\Bar{L}$ is an approximate of the eigenvalues of $\bm{L}$, such as
\begin{align}
    \bar{L}&=\left(\frac{1}{M}\mathrm{trace}\left(\bm{L}^\alpha\right)\right)^{\frac{1}{\alpha}},\label{eq:L_bar}
\end{align}
in which $\alpha$ is a design parameter.
\end{lemma}
\begin{proof}
The covariance matrix $\bm{L}$ has the eigenvalue decomposition $\bm{L}=\bm{W}\mathrm{diag}\left(L_1,\dots,L_M\right)\bm{W}^H$. From this, the values $l_m=\left(\bm{W}^H\bm{l}\right)_m$ are distributed as $l_m\sim\mathcal{CN}\left(\underline{l}_m,L_{m}\right)$, in which $\underline{l}_m=\left(\bm{W}^H\bm{\underline{l}}\right)_m$. By approximating the variance $L_{m}$ with the value of \eqref{eq:L_bar}, the real and imaginary parts of all $\sqrt{\frac{2}{\bar{L}}}l_m$ are real-valued Gaussian distributed with unit variance. Hence, the sum of the squares of these variables is distributed as non-central chi-square distribution with $2M$ degrees of freedom, i.e.,
\begin{align}
    \frac{2}{\bar{L}}\sum_{m=1}^M l_m^2\sim\chi^2\left(2M,\frac{2}{\bar{L}}\sum_{m=1}^M\left|\underline{l}_m\right|^2\right).
\end{align}
For this, the cumulative distribution function (CDF) can be written on behalf of the Marcum $Q$-function, i.e., \cite{1055327}
\begin{align}
    F(x)=1-Q_M\left(\sqrt{\frac{2}{\bar{L}}\sum_{m=1}^M\left|\underline{l}_m\right|^2},\sqrt{x}\right).\label{eq:MarcumQdef}
\end{align}
Evaluating \eqref{eq:MarcumQdef} at $x=\frac{2\eta}{\bar{L}}$, we get \eqref{eq:cdf_value}.
\end{proof}
As the channel measurement and process measurement are uncorrelated, the number of true and false negatives can be obtained for each time instant $t$ as
\begin{align}
&P_{\{\mathrm{TN},\mathrm{FN}\}}(t)\nonumber\\&\hspace{0.025cm}=\left(1-Q_{M_{\mathrm{P}}}\left(\sqrt{\frac{2}{\bar{L}_{\mathrm{P}}}\bm{\underline{e}}_{\mathrm{P},g}^H(t)\bm{V}_{\mathrm{P},g}\bm{\underline{e}}_{\mathrm{P},g}(t)},\sqrt{\frac{2\eta_{\mathrm{P}}}{\bar{L}_{\mathrm{P}}}}\right)\right)\nonumber\\&\hspace{0.05cm}\times\left(1-Q_{M_{\mathrm{C}}}\left(\sqrt{\frac{2}{\bar{L}_{\mathrm{C}}}\bm{\underline{e}}_{\mathrm{C},g}^H(t)\bm{V}_{\mathrm{C},g}\bm{\underline{e}}_{\mathrm{C},g}(t)},\sqrt{\frac{2\eta_{\mathrm{C}}}{\bar{L}_{\mathrm{C}}}}\right)\right),\label{eq:p_marcum}
\end{align}
where $\bm{\underline{e}}_{\mathrm{P},g}(t)=\mathds{E}\left[\bm{e}_{\mathrm{P},g}(t)\right]$ and $\bar{L}_{\mathrm{P}}=\frac{1}{M_{\mathrm{P}}}\mathrm{trace}\left(\bm{V}_{\mathrm{P},g}\mathds{E}\left[\bm{e}_{\mathrm{P},g}(t)\bm{e}_{\mathrm{P},g}^H(t)\right]\right)$; $\bm{\underline{e}}_{\mathrm{C},g}(t)$ and $\bar{L}_{\mathrm{C}}$ are defined similarly. For the legitimate and adversarial transmitter, the former can be derived as
\begin{align}
    \bm{\underline{e}}_{\mathrm{P},g}(t)&=\begin{cases}\bm{0}&\text{legitimate transmitter}\\\bm{\tilde{\underline{y}}}_{\mathrm{P},g}(t)-\bm{C}_{\mathrm{P},g}\bm{\hat{\underline{z}}}_{\mathrm{P}}(t|t_{\mathrm{P}})\hspace{-1.3cm}\ \\&\text{adversarial transmitter}\end{cases}.\label{eq:e_p_options}
\end{align}
For $\bar{L}_{\mathrm{P}}$, we get in the two cases
\begin{align}
    &\bar{L}_{\mathrm{P}}\nonumber\\&=\begin{cases}\left(\frac{1}{M_{\mathrm{P}}}\mathrm{trace}\left(\left(\bm{V}_{\mathrm{P},g}\left(\bm{C}_{\mathrm{P},g}\bm{\hat{Z}}(t|t_{\mathrm{P}})\bm{C}_{\mathrm{P},g}^H+\bm{N}_{\mathrm{P}}\right)\right)^\alpha\right)\right)^{\frac{1}{\alpha}}\hspace{-3.7cm}\ \\&\text{legitimate transmitter}\\\left(\frac{1}{M_{\mathrm{P}}}\mathrm{trace}\left(\left(\bm{V}_{\mathrm{P},g}\bm{\tilde{Y}}_{\mathrm{P},g}(t)\right)^\alpha\right)\right)^{\frac{1}{\alpha}}\hspace{-1.2cm}\ &\\&\text{adversarial transmitter}\end{cases}.\label{eq:l_p_options}
\end{align}
For the legitimate transmitter, the Marcum $Q$-function in \eqref{eq:p_marcum} can be simplified to the regularized incomplete gamma function due to \eqref{eq:e_p_options}. Hence, the ratio of true negatives can be written as
\begin{align}
    P_{\mathrm{TN}}(t)=Q\left(M_{\mathrm{P}},\frac{\eta_{\mathrm{P}}}{\bar{L}_{\mathrm{P}}}\right)Q\left(M_{\mathrm{C}},\frac{\eta_{\mathrm{C}}}{\bar{L}_{\mathrm{C}}}\right).\label{eq:p_gamma}
\end{align}
For a chosen pair of threshold values, equations \eqref{eq:p_marcum} and \eqref{eq:p_gamma} can be used to approximate the probabilities of false and true negatives at each timestamp. To limit the adversary's success probability to a specific tolerated value, the Marcum $Q$-function can be inverted, which has been studied in \cite{https://doi.org/10.1111/sapm.12050}. As $Q_M(x,y)$ is monotonically decreasing in $y$ and has values in $[0,1]$, \eqref{eq:p_marcum} is monotonically increasing in $\eta_{\mathrm{P}}$ and $\eta_{\mathrm{C}}$. 

With the malicious sensor node and the adversarial devices positioned in the environment, there are two different kinds of attackers, of which each has an own success probability $P_{\mathrm{FN}}(t)$. To indicate that a success rate belongs to any of the two attackers, we add an index $(1)$ or $(2)$ to the corresponding variable. If the target is to limit a tolerated success probability of the adversaries to $P_{\mathrm{FN}}^\star$ for each received packet, the threshold vector $\bm{\eta}=(\eta_{\mathrm{P}},\eta_{\mathrm{C}})^T$ can be optimized algorithmically. Therefore, an initial $\bm{\eta}_0$ can be updated iteratively by applying
\begin{align}
    \bm{\eta}_{n+1}=\bm{\eta}_n-\bm{B}\begin{pmatrix}P_{\mathrm{FN},n}^{(1)}(t)-P_{\mathrm{FN}}^\star\\P_{\mathrm{FN},n}^{(2)}(t)-P_{\mathrm{FN}}^\star\end{pmatrix}\label{eq:gradient_for_eta}
\end{align}
until convergence, where $\bm{B}$ represents the step width. For the first received packet $\bm{\eta}_{0}$ is chosen arbitrarily. For follow-up packets, $\bm{\eta}_{0}$ can be initiated as the result optimized at the previous timestamp.

\subsection{Gaussian Approximation for large $M_{\mathrm{P}}$ and $M_{\mathrm{C}}$}\label{sec:Gauss}

In some wireless sensor networks, the fusion center is connected to a base station, which is supporting massive SIMO techniques. Also, the sensors might measure various information together, as in case of an image sensor. When $M_{\mathrm{P}}$ and $M_{\mathrm{C}}$ are large, the probability of remaining below the threshold is analyzed for a simplified statement in Lemma~\ref{lem:gaussian}.

\begin{lemma}\label{lem:gaussian}
If $\bm{l}\sim\mathcal{CN}\left(\bm{\underline{l}},\bm{L}\right)$
is a stochastic variable, the value $s=\bm{l}^H\bm{l}$ can be approximated by the Gaussian
\begin{align}
    s\sim\mathcal{N}\left(\underline{s},S\right),
\end{align}
of which the mean and covariance matrix are
\begin{align}
    \underline{s}&=\mathrm{trace}\left(\bm{L}\right)+\bm{\underline{l}}^H\bm{\underline{l}}\label{eq:eta_mean}\\
    S&=\mathrm{trace}\left(\bm{L}^2\right)+2\bm{\underline{l}}^H\bm{L}\bm{\underline{l}},\label{eq:eta_var}
\end{align}
respectively. Hence, the statement $s\leq\eta$ holds true with probability
\begin{align}
    P=Q\left(-\frac{\eta-\underline{s}}{\sqrt{S}}\right).\label{eq:sleqetaGauss}
\end{align}
\end{lemma}
\begin{proof}
The covariance matrix $\bm{L}$ has the eigenvalue decomposition $\bm{L}=\bm{W}\mathrm{diag}\left(L_1,\dots,L_M\right)\bm{W}^H$. From this, the values $l_m=\left(\bm{W}^H\bm{l}\right)_m$ are distributed as $l_m\sim\mathcal{CN}\left(\underline{l}_m,L_m\right)$, in which $\underline{l}_m=\left(\bm{W}^H\bm{\underline{l}}\right)_m$.
In the simplified case of a real-valued $l\sim\mathcal{N}(\underline{l},1)$, $\lambda=l^2$ is distributed as
\begin{align}
    \lambda\sim\begin{cases}
    \frac{\exp\left(-\frac{1}{2}\left(\sqrt{\lambda}-\underline{l}\right)^2\right)+\exp\left(-\frac{1}{2}\left(\sqrt{\lambda}+\underline{l}\right)^2\right)}{\sqrt{8\pi\lambda}}&\lambda\geq 0\\
    0& \lambda<0
    \end{cases}.\label{eq:chi_square}
\end{align}
The mean and variance of $\lambda$ are $\underline{\lambda}=1+\underline{l}^2$ and $\Lambda=2+4\underline{l}^2$, respectively. If the variance is generalized, i.e., $l_m^o\sim\mathcal{N}(\underline{l}_m^o,\frac{1}{2}L_m)$, we can substitute $\lambda$ by $\frac{1}{2}L_m\lambda_m^o$ and $\underline{l}$ by $2\underline{l}_m^o/\sqrt{\lambda_i^o}$, to derive that $\lambda_m^o=(l_m^o)^2$ has mean $\underline{\lambda}_m^o=\frac{1}{2}L_i+(\underline{l}_m^o)^2$ and variance $\Lambda_m=\frac{1}{2}L_m^2+2(\underline{l}_m^o)^2L_m$. When calculating the sum over all $m\in\{1,\dots,M\}$ and $o\in\{\mathrm{R},\mathrm{I}\}$ with $\underline{l}_m^{\mathrm{R}}$ and $\underline{l}_m^{\mathrm{I}}$ being the real and imaginary part of $\underline{l}_m$, we get
\begin{align}
    \underline{s}&=\sum_{m=1}^M L_m+\sum_{m=1}^M\left|\underline{l}_m\right|^2\label{eq:s_mean}\\
    S&=\sum_{m=1}^M L_m^2+2\sum_{m=1}^M\left|\underline{l}_m\right|^2L_m,\label{eq:s_var}
\end{align}
which are equal to \eqref{eq:eta_mean} and \eqref{eq:eta_var}.
\end{proof}

Using lemma~\ref{lem:gaussian} and \eqref{eq:e_p_options}, we can approximate $s_{\mathrm{P}}(t)=\bm{e}_{\mathrm{P},g}^H(t)\bm{V}_{\mathrm{P},g}\bm{e}_{\mathrm{P},g}(t)$ for the legitimate transmitter as Gaussian with mean and variance
\begin{align}
    \underline{s}_{\mathrm{P}}(t)&=\mathrm{trace}\left\{\bm{V}_{\mathrm{P},g}\left(\bm{C}_{\mathrm{P},g}\bm{\hat{Z}}_{\mathrm{P}}(t|t')\bm{C}_{\mathrm{P},g}^H+\bm{N}_{\mathrm{P}}\right)\right\},\\
    S_{\mathrm{P}}(t)&=\mathrm{trace}\left\{\left(\bm{V}_{\mathrm{P},g}\left(\bm{C}_{\mathrm{P},g}\bm{\hat{Z}}_{\mathrm{P}}(t|t')\bm{C}_{\mathrm{P},g}^H+\bm{N}_{\mathrm{P}}\right)\right)^2\right\}.
\end{align}
For the adversary, the same expression is distributed as Gaussian with parameters
\begin{align}
    \tilde{\underline{s}}_{\mathrm{P}}(t)&=\mathrm{trace}\left\{\bm{V}_{\mathrm{P},g}\bm{\tilde{Y}}_{\mathrm{P}}(t)\right\}+\left(\bm{C}_{\mathrm{P},g}\bm{\hat{\underline{z}}}_{\mathrm{P}}(t|t')-\bm{\tilde{\underline{y}}}_{\mathrm{P},g}(t)\right)^H\nonumber\\&\hspace{0.6cm}\times\bm{V}_{\mathrm{P}}\left(\bm{C}_{\mathrm{P},g}\bm{\hat{\underline{z}}}_{\mathrm{P}}(t|t')-\bm{\tilde{\underline{y}}}_{\mathrm{P},g}(t)\right)\\
    \tilde{S}_{\mathrm{P}}(t)&=\mathrm{trace}\left\{\left(\bm{V}_{\mathrm{P},g}\bm{\tilde{Y}}_{\mathrm{P}}(t)\right)^2\right\}\nonumber\\&\hphantom{=}+2\left(\bm{C}_{\mathrm{P},g}\bm{\hat{\underline{z}}}_{\mathrm{P}}(t|t')-\bm{\tilde{\underline{y}}}_{\mathrm{P},g}(t)\right)^H\bm{V}_{\mathrm{P},g}\bm{\tilde{Y}}_{\mathrm{P}}(t)\bm{V}_{\mathrm{P},g}\nonumber\\&\hspace{0.6cm}\times\left(\bm{C}_{\mathrm{P},g}\bm{\hat{\underline{z}}}_{\mathrm{P}}(t|t')-\bm{\tilde{\underline{y}}}_{\mathrm{P},g}(t)\right).
\end{align}
Similar expressions can be obtained for the channel-related parameters. The number of true negatives can be obtained by integrating the approximated probability density function (PDF) of the legitimate transmitter from the left towards the thresholds $\eta_{\mathrm{P}}$ and $\eta_{\mathrm{C}}$. This value becomes
\begin{align}
    P_{\mathrm{TN}}(t)=Q\left(-\frac{\eta_{\mathrm{P}}-\underline{s}_{\mathrm{P}}(t)}{\sqrt{S_{\mathrm{P}}(t)}}\right)Q\left(-\frac{\eta_{\mathrm{C}}-\underline{s}_{\mathrm{C}}(t)}{\sqrt{S_{\mathrm{C}}(t)}}\right).
\end{align}
A similar expression holds for the number of false negatives
\begin{align}
    P_{\mathrm{FN}}(t)=Q\left(-\frac{\eta_{\mathrm{P}}-\tilde{\underline{s}}_{\mathrm{P}}(t)}{\sqrt{\tilde{S}_{\mathrm{P}}(t)}}\right)Q\left(-\frac{\eta_{\mathrm{C}}-\tilde{\underline{s}}_{\mathrm{C}}(t)}{\sqrt{\tilde{S}_{\mathrm{C}}(t)}}\right).\label{eq:P_FN_Gauss}
\end{align}

As there are two types of attackers, i.e., malicious sensor nodes and adversarial devices positioned in the environment, there are two different sets of parameters, and thus also two different values of $P_{\mathrm{FN}}(t)$ will be achieved. To indicate that a specific variable belongs to any of the two sets of parameters, we add an index $(1)$ or $(2)$ to the corresponding variables. As the target is to fulfill a certain security constraint in the form of a tolerated $P_{\mathrm{FN}}^\star$ for each received packet, the threshold values $\eta_{\mathrm{P}}$ and $\eta_{\mathrm{C}}$ can be obtained iteratively by calculating
\begin{align}
    \eta_{\mathrm{P},n+1}&=-\sqrt{\tilde{S}_{\mathrm{P}}^{(1)}(t)}Q^{-1}\left(\frac{P_{\mathrm{FN}}^\star}{P_{\mathrm{C},n}^{(1)}}\right)+\tilde{\underline{s}}_{\mathrm{P}}^{(1)}(t)\label{eq:eta_P_choice}\\\eta_{\mathrm{C},n+1}&=-\sqrt{\tilde{S}_{\mathrm{C}}^{(2)}(t)}Q^{-1}\left(\frac{P_{\mathrm{FN}}^\star}{P_{\mathrm{P},n}^{(2)}}\right)+\tilde{\underline{s}}_{\mathrm{C}}^{(2)}(t)\label{eq:eta_C_choice}
\end{align}
until convergence. Thereby, $P_{\mathrm{C},0}^{(1)}$ and $P_{\mathrm{P},0}^{(2)}$ are initiated to one. For each calculated pair of threshold values, $P_{\mathrm{C},n}^{(1)}$ and $P_{\mathrm{P},n}^{(2)}$ become the second and first part of \eqref{eq:P_FN_Gauss} for the first and second attacker model, respectively.

\subsection{Approximation for asymptotically large $M_{\mathrm{P}}$ and $M_{\mathrm{C}}$}
In massive SIMO systems with fading channels, channel hardening becomes relevant \cite{7880691}. This means that the stochastic effect of the channel from the sensors to each base station antenna averages out when the number of antennas at the base station becomes large. This behavior is equivalent to the case of asymptotically large output vectors, i.e., very large $M_{\mathrm{P}}$ and $M_{\mathrm{C}}$. This means that the entries of the prediction vector $\bm{\hat{\underline{z}}}_{\mathrm{P}}(t|t_{\mathrm{P}})$ obtained by the Kalman filter can themselves be seen as a Gaussian distributed variable. From this, also the entries of the expectation vector $\bm{\underline{e}}_{\mathrm{P},g}(t)$ in \eqref{eq:e_p_options} become Gaussian distributed. In case of the legitimate transmitter, the mean and the variance are zero due to \eqref{eq:e_p_options}. For the case that the adversarial transmitter is active, $\bm{u}_{\mathrm{P},g}(t)$ is zero-mean and all components are linear. Hence, we conclude that $\bm{\hat{\underline{z}}}_{\mathrm{P}}(t|t_{\mathrm{P}})$ is zero-mean. We assume the same holds also true for $\bm{\tilde{\underline{y}}}_{\mathrm{P},g}(t)$, as the adversarial measurements are either distributed similarly to the measurements of the legitimate transmitter or the adversary is modifying their measurements in an unbiased way.

To analyze the variance of the mean $\bm{\underline{e}}_{\mathrm{C},g}(t)$ in the case of the adversarial transmission, we need to derive the variances of the Kalman filter estimates, i.e., the variances of the estimates in equations \eqref{eq:z_mean_init}, \eqref{eq:e_P}, \eqref{eq:Kalman2}, and \eqref{eq:Kalman4}. Obviously, the constant in \eqref{eq:z_mean_init} has a covariance of $\bm{\hat{\underline{Z}}}_{\mathrm{P}}(t|-1)=\bm{0}$. Moreover, the innovation \eqref{eq:e_P} has a variance of
\begin{align}
    \bm{E}_{\mathrm{P},g}(t)=\bm{C}_{\mathrm{P},g}\bm{\hat{Z}}_{\mathrm{P}}(t|t_{\mathrm{P}})\bm{C}_{\mathrm{P},g}^H+\bm{N}_{\mathrm{P}}.
\end{align}
From this, the variance of the state expectation from the Kalman filter in \eqref{eq:Kalman2} is
\begin{align}
    \bm{\hat{\underline{Z}}}_{\mathrm{P}}(t)=\bm{\hat{\underline{Z}}}_{\mathrm{P}}(t|t_{\mathrm{P}})+\bm{K}_{\mathrm{P}}(t)\bm{E}_{\mathrm{P},g}(t)\bm{K}_{\mathrm{P}}^H(t).
\end{align}
This value is further used to obtain the variance of the expectation of the prediction in \eqref{eq:Kalman4} as
\begin{align}
    \bm{\hat{\underline{Z}}}_{\mathrm{P}}(t|t_{\mathrm{P}})=e^{\bm{A}_{\mathrm{P}}(t-t_{\mathrm{P}})}\bm{\hat{\underline{Z}}}_{\mathrm{P}}(t_{\mathrm{P}})e^{\bm{A}_{\mathrm{P}}^H(t-t_{\mathrm{P}})}.
\end{align}
To be able to calculate the variance of the expected innovation, we need to model the dependency of the expectations of the process value from the adversary and the Kalman filter prediction. Therefore, we assume that $\bm{\tilde{\underline{y}}}_{\mathrm{P},g}(t)=\bm{\tilde{C}}_{\mathrm{P},g}\bm{\hat{\underline{z}}}_{\mathrm{P}}(t|t_{\mathrm{P}})$, in which $\bm{\tilde{C}}_{\mathrm{P},g}=\bm{0}$ if the adversary has zero-knowledge and $\bm{\tilde{C}}_{\mathrm{P},g}=\bm{C}_{\mathrm{P},g}$ if the adversary has full knowledge about the measurement from the process system. From this, the entries of the expectation vector are distributed as
\begin{align}
    \bm{\underline{e}}_{\mathrm{P},g}(t)\sim\mathcal{CN}\left(\bm{0},\bm{\underline{E}}_{\mathrm{P},g}(t)\right),
\end{align}
where 
\begin{align}
    \bm{\underline{E}}_{\mathrm{P},g}(t)=\left(\bm{\tilde{C}}_{\mathrm{P},g}-\bm{C}_{\mathrm{P},g}\right)\bm{\hat{\underline{Z}}}_{\mathrm{P}}(t|t_{\mathrm{P}})\left(\bm{\tilde{C}}_{\mathrm{P},g}^H-\bm{C}_{\mathrm{P},g}^H\right).
\end{align}
To analyze the results of the asymptotic case, Lemma~\ref{lem:asymptotic} is used. 
\begin{lemma}\label{lem:asymptotic}
If $\bm{\underline{l}}\sim\mathcal{CN}\left(\bm{0},\bm{\underline{L}}\right)$ and $\bm{l}\sim\mathcal{CN}\left(\bm{\underline{l}},\bm{L}\right)$ are stochastic variables, the variable $\bm{s}=\bm{l}^H\bm{l}$ can be approximated by the Gaussian
\begin{align}
    s\sim\mathcal{N}\left(\mathrm{trace}\left(\bm{L}+\bm{\underline{L}}\right),\mathrm{trace}\left(\left(\bm{L}+\bm{\underline{L}}\right)^2\right)\right).
\end{align}
Hence, $s\leq\eta$ is fulfilled with a probability similar to \eqref{eq:sleqetaGauss}.
\end{lemma}
\begin{proof}
As $\bm{l}=\bm{\underline{l}}+\bm{l'}$ and $\bm{l'}\sim\mathcal{CN}\left(0,\bm{L}\right)$, $\bm{l}\sim\mathcal{CN}\left(0,\bm{L}+\bm{\underline{L}}\right)$. For this variable, Lemma~\ref{lem:gaussian} holds.
\end{proof}
Employing this, $\tilde{s}_{\mathrm{P}}(t)=\bm{e}_{\mathrm{P},g}^H(t)\bm{V}_{\mathrm{P},g}\bm{e}_{\mathrm{P},g}(t)$ can be approximated for the adversarial transmitter as Gaussian distributed, where the mean and variance are
\begin{align}
    \tilde{\underline{s}}_{\mathrm{P}}(t)&=\mathrm{trace}\left\{\bm{V}_{\mathrm{P},g}\left(\bm{\tilde{Y}}_{\mathrm{P}}(t)+\bm{\underline{E}}_{\mathrm{P},g}(t)\right)\right\}\\
    \tilde{S}_{\mathrm{P}}(t)&=\mathrm{trace}\left\{\left(\bm{V}_{\mathrm{P},g}\left(\bm{\tilde{Y}}_{\mathrm{P}}(t)+\bm{\underline{E}}_{\mathrm{P},g}(t)\right)\right)^2\right\}.
\end{align}
Employing these parameters, the number of false negatives can be obtained by \eqref{eq:P_FN_Gauss}, while the number of true negatives equals the result of the previous section. Similarly, the optimal threshold values can be found in \eqref{eq:eta_P_choice} and \eqref{eq:eta_C_choice}.

\section{Numerical Results}
We consider a system with $G$ sensors, which are scheduled by a round-robin scheduling scheme to transmit their sensor data to the fusion center. Thereby, the $g$-th sensor is scheduled to transmit measurements at time instants $g+nG$, in which $n\in\mathds{N}$. While the first $G$ time instants are used for the login phases of the different sensors, the following time instants are used for authentication. Through the numerical results, we are choosing the weight matrices as \cite{7470566}
\begin{align}
    \bm{V}_{\mathrm{P},g}&=\left(\bm{C}_{\mathrm{P},g}\bm{\hat{Z}}_{\mathrm{P}}(t|t_{\mathrm{P}})\bm{C}_{\mathrm{P},g}^H+\bm{N}_{\mathrm{P},g}\right)^{-1},\\
    \bm{V}_{\mathrm{C},g}&=\left(\bm{C}_{\mathrm{C},g}\bm{\hat{Z}}_{\mathrm{C},g}(t|t_{\mathrm{C},g})\bm{C}_{\mathrm{C},g}^H+\bm{N}_{\mathrm{C},g}\right)^{-1}.
\end{align}
Next, we will investigate the impact of the threshold choice on the security and the reliability of a system. Afterward, we will analyze the security and reliability over different channel vector lengths.

\begin{figure}
    \centering
    \subfloat[True negatives]{\includegraphics{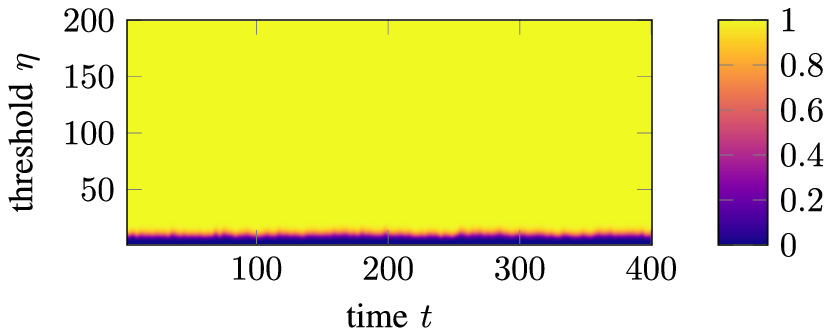}}\hfill
    \subfloat[False negatives for a malicious sensor node]{\includegraphics{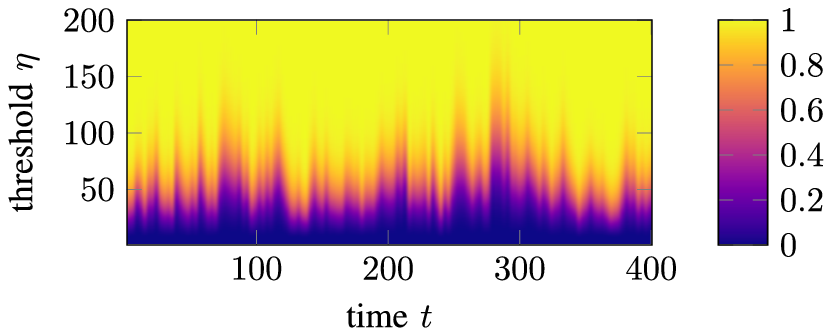}}\hfill
    \subfloat[False negatives for an adversarial device]{\includegraphics{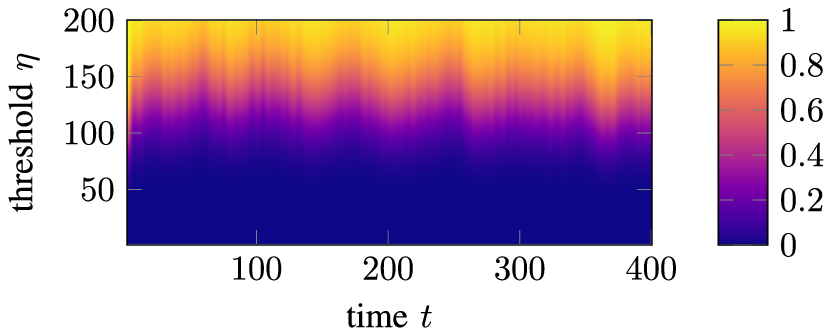}}
    \caption{The impact of the threshold choice on reliability (i.e., the number of true negatives) and security (i.e., the number of false negatives).}
    \label{fig:variousEta}
\end{figure}

\subsection{Impact of the Threshold Choice}

The analytical results indicate that the threshold choice has a crucial impact on the reliability and security of the system.
To investigate this impact, we simulate a system with $G=3$ sensors, which transmit packets over a sequence of $400$ time instants. At each time instant, we generate $\num{100000}$ possible follow-up packets from each transmitter setting and analyze the averaged behavior over these packets. Afterward, we assume one legitimate packet to be authenticated, before the next packet is received.

The considered process system has the following parameters. The system matrix $\bm{A}_{\mathrm{P}}=-\frac{0.2}{G}\mathrm{diag}\left(\bm{a}_{\mathrm{P}}\right)$ is chosen as a $10\times10$-matrix, where the entries of $\bm{a}_{\mathrm{P}}$ are equally distributed on $[0.5,1]$. Moreover, the input noise is specified by $\bm{U}_{\mathrm{P}}=\frac{0.2}{G}\bm{\ddot{U}}_{\mathrm{P}}\bm{\ddot{U}}_{\mathrm{P}}^H$, in which the entries of $\bm{\ddot{U}}_{\mathrm{P}}$ are distributed equally within $[-1,1]$. The output matrices $\bm{C}_{\mathrm{P},g}$ are matrices of size $4\times 10$, whose entries are uniformly distributed within $[0,1]$. Finally, the covariance matrices of the measurement noise are $\bm{N}_{\mathrm{P}}=0.01\bm{\ddot{N}}_{\mathrm{P}}\bm{\ddot{N}}_{\mathrm{P}}^H$, in which $\bm{\ddot{N}}_{\mathrm{P}}$ is distributed similarly to $\bm{\ddot{U}}_{\mathrm{P}}$.

For the channel model, we employ the parameters from \cite{9034987} with $M_{\mathrm{C}}=10$. Thereby, the system matrix is specified by the $M_{\mathrm{C}}\times M_{\mathrm{C}}$-matrix $\bm{A}_{\mathrm{C}}=a_{\mathrm{C}}\bm{I}$, in which $a_{\mathrm{C}}=-\log(2)\times10^{-2}/G$. The covariance matrix of the input noise is $\bm{U}_{\mathrm{C}}=-2a_{\mathrm{C}}\bm{I}$, while the output matrices are $\bm{C}_{\mathrm{C},g}=\bm{I}$, which have both the same dimensions as $\bm{A}_{\mathrm{C}}$. Here, the measurement noise is chosen such that the SNR is $10\,\mathrm{dB}$.

We first simulate for various threshold values of $\eta=\eta_{\mathrm{P}}=\eta_{\mathrm{C}}$ the reliability, i.e., the number of true negatives for the legitimate transmitter, and the security, i.e., the number of false negatives for both kinds of attackers. As all $G$ sensors follow a similar behavior, we focus on the first sensor and show the corresponding numerical results in \figurename~\ref{fig:variousEta}. First, the results show that the equal choice of the two threshold values $\eta_{\mathrm{P}}$ and $\eta_{\mathrm{C}}$ has a very different impact on the two kinds of attackers as the parameters of the process system and the channel system differ. Moreover, the results show that for a fixed $\eta$, the security level is time-varying. This shows that it is required to choose time-varying threshold values if the security level should be kept constant over time.

\begin{figure}
    \centering
    \subfloat[Threshold value $\eta_\mathrm{P}$ over time]{\includegraphics{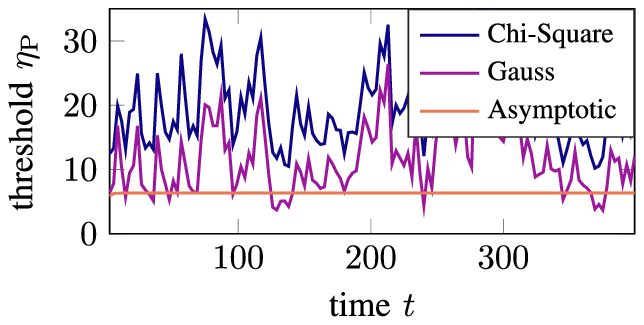}}\hfill
    \subfloat[Threshold value $\eta_\mathrm{C}$ over time]{\includegraphics{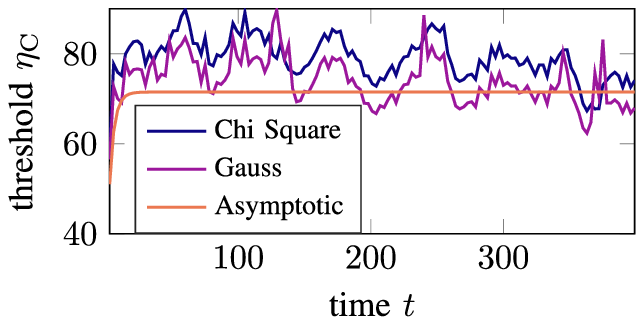}}
    \caption{The optimized threshold values for the three methods provided.}
    \label{fig:optimizedThreshold}
\end{figure}

\begin{figure}
    \centering
    \subfloat[True negatives]{\includegraphics{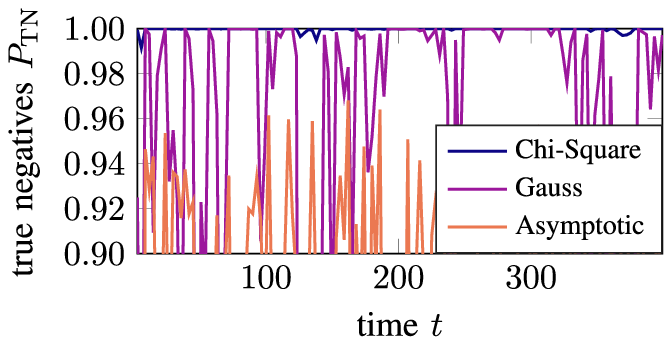}}\hfill
    \subfloat[False negatives for a malicious sensor node]{\includegraphics{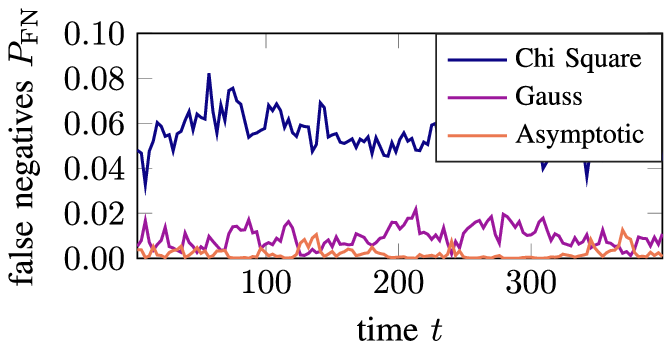}}\hfill
    \subfloat[False negatives for an adversarial device]{\includegraphics{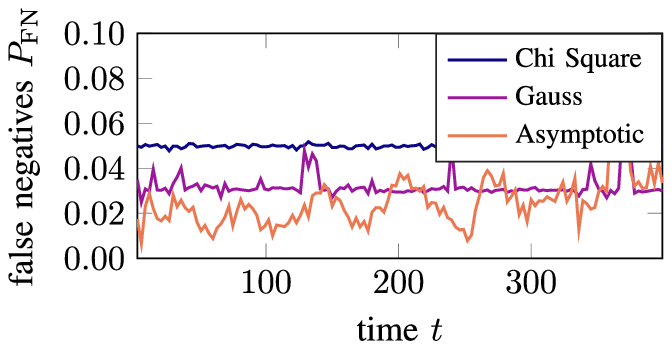}}
    \caption{The impact of the optimized threshold values on reliability (i.e., the number of true negatives) and security (i.e., the number of false negatives) for the three methods provided.}
    \label{fig:optimizedEta}
\end{figure}

Next, we fix the tolerated number of false negatives to $P_{\mathrm{FN}}^\star=0.05$ and optimize the pairs of threshold values for each time instant. We focus again on the first of the three sensors and show the optimized threshold values in \figurename~\ref{fig:optimizedThreshold}. The results show that the threshold values obtained with the chi-square approximation (with $\alpha=0.5$) and the Gaussian approximation show similar temporal behavior. For the two methods, only the amplitudes of the threshold values chosen slightly differ. 
While the threshold choice of the chi-square and Gaussian approximation show a significant temporal dependency, the temporal dependency of the thresholds obtained with the asymptotic approximation is reduced. 
Thereby, the optimized threshold values are often below the values obtained with the other methods. Moreover, the asymptotic approximation shows that there is a transient process occurring at the beginning until the system behavior reaches a stabilized functionality. To analyze the quality of the threshold choices, we now simulate the temporal behavior of the numbers of true and false negatives, which can be obtained for the selected threshold choices. \figurename~\ref{fig:optimizedEta} shows that in all cases the numbers of false negatives fluctuate around different constant values. In the simulation considered here, the numbers of false negatives $P_{\mathrm{FN}}$ of the chi-square approximation match the tolerated $P_{\mathrm{FN}}^\star$ best, but this might vary for differently chosen parameters. However, due to the temporal dependency of $\eta_{\mathrm{P}}$ and $\eta_{\mathrm{C}}$, larger fluctuations occur for the simulated number of true negatives (especially in the case of the Gaussian approximation). This means that for a small fraction of the packets, only relatively low reliability can be obtained. For these packets, one option is to introduce additional methods to secure these packets with a higher computational complexity at a cost. However, the results also show that this effect can be limited by using a larger threshold value, such as the one of the chi-square approximation, as long as the security constraint remains fulfilled. From these results, it remains open for now, how the exact system performance changes over $M_{\mathrm{P}}$ and $M_{\mathrm{C}}$. We address this in the next subsection.

\begin{figure*}
    \centering
    \subfloat[Average true negatives over $M_{\mathrm{C}}$]{\includegraphics{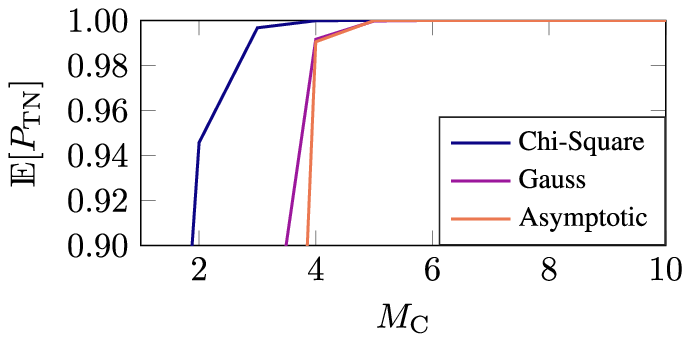}} 
    \subfloat[Average false negatives over $M_{\mathrm{C}}$\label{fig:pOverM-EPFN}]{\includegraphics{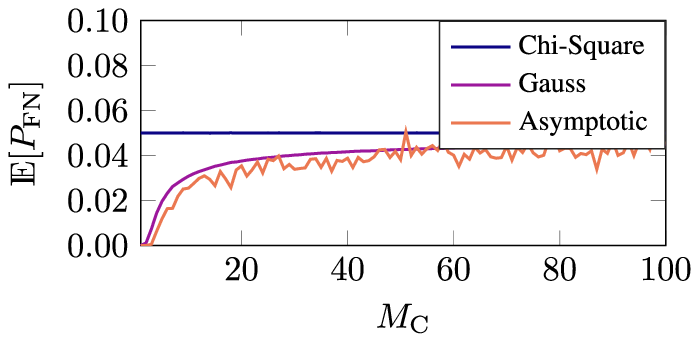}}\\
    \subfloat[CDF of the false negatives  $M_{\mathrm{C}}=10$\label{fig:pOverM-FN10}]{\includegraphics{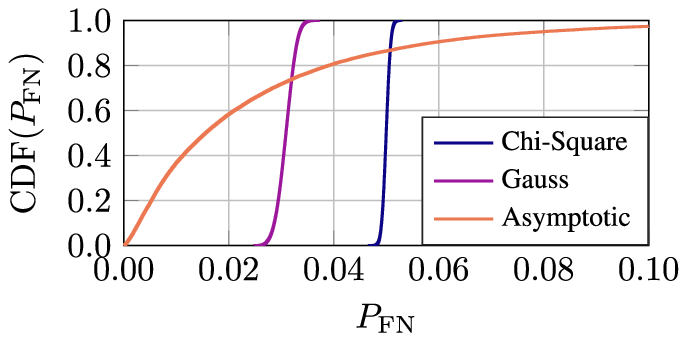}} 
    \subfloat[CDF of the false negatives  $M_{\mathrm{C}}=100$\label{fig:pOverM-FN100}]{\includegraphics{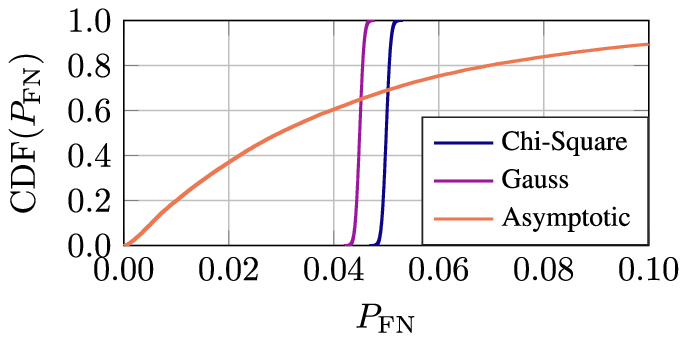}}
    \caption{Empirical expectation of $P_{\mathrm{TN}}$ and $P_{\mathrm{FN}}$ over $M_{\mathrm{C}}$ and CDF of $P_{\mathrm{FN}}$ for two values of $M_{\mathrm{C}}$ with $P_{\mathrm{FN}}^\star=0.05$ with an identity-based matrix structure for the three methods provided.}
    \label{fig:pOverM}
\end{figure*}

\subsection{Impact of $M_{\mathrm{P}}$ and $M_{\mathrm{C}}$}
The theoretical analysis has shown that the quality of the approximations depends on the vector lengths $M_{\mathrm{P}}$ and $M_{\mathrm{C}}$. To investigate the system behavior over these parameters, we focus on a system with $G=1$ sensor, in which only one of the two kinds of measurements is employed for authentication (i.e., by setting $M_{\mathrm{P}}=0$). We simulate the probability of true and false negatives for the packets received at each time instant similar to the previous section. The results are then used to empirically calculate the mean and the CDF of these probabilities based on packets received over $400$ time instants and $100$ transmission sequences. Note that at each time instant, the probabilities are generated numerically by investigating the number of true and false negatives occurring over a constant number of packets. Hence, the simulation introduces a binomial distribution for each received packet, which reduces the steepness of the CDF curves. To limit this effect, we use $\num{100000}$ generated follow-up packets for the CDF plots. When calculating the mean of the probabilities, $2000$ possible follow-up packets are sufficient.

Thereby, we first focus on the case of the channel with identity-based parameters from \cite{9034987} as presented above. 
\figurename~\ref{fig:pOverM} shows the mean of the two probabilities over the vector length $M_{\mathrm{C}}$ and the CDF of $P_{\mathrm{FN}}$ for two values of $M_{\mathrm{C}}$. In the considered special case of scaled identity matrices, no approximations are involved in the chi-square plots. Consequently, \figurename~\ref{fig:pOverM-EPFN} shows that the chi-square approximation allows an excellent fulfillment of the tolerated $P_{\mathrm{FN}}^\star$. As the quality of the Gaussian approximation increases over $M_\mathrm{C}$, the corresponding expected false alarm rate converges to $P_{\mathrm{FN}}^\star$ for large $M_\mathrm{C}$. \figurename{S}~\ref{fig:pOverM-FN10} and~\ref{fig:pOverM-FN100} show that both approximations lead to a security level, which is almost constant over time. Similar to the mean, also the CDF of the number of false negatives obtained with the Gaussian approximation converges towards the CDF of the chi-square approximation for large $M_{\mathrm{C}}$. The asymptotic approximation however requires a longer vector length until the CDF convergences.

\begin{figure*}
    \centering
    \subfloat[Average true negatives over $M_{\mathrm{C}}$]{\includegraphics{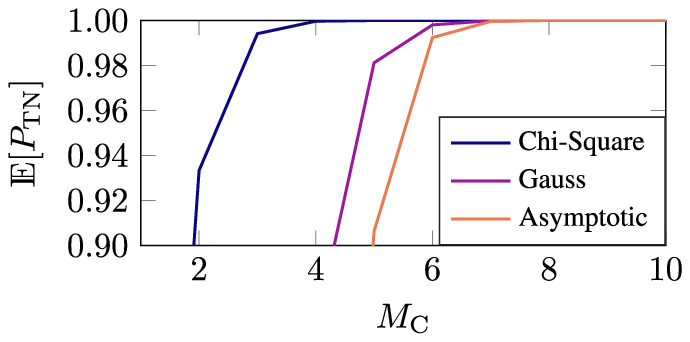}} 
    \subfloat[Average false negatives over $M_{\mathrm{C}}$\label{fig:pOverM-EPFN2}]{\includegraphics{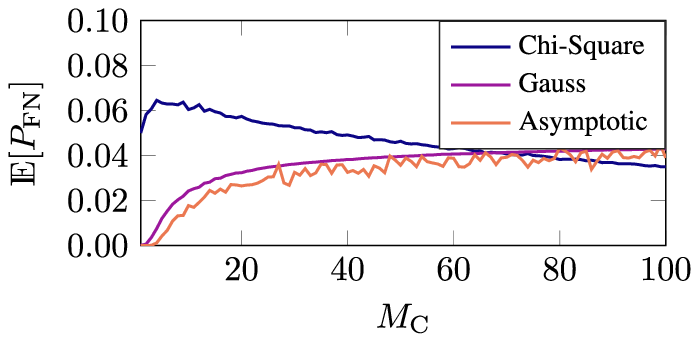}}\\
    \subfloat[CDF of the false negatives  $M_{\mathrm{C}}=10$\label{fig:pOverM2-FN10}]{\includegraphics{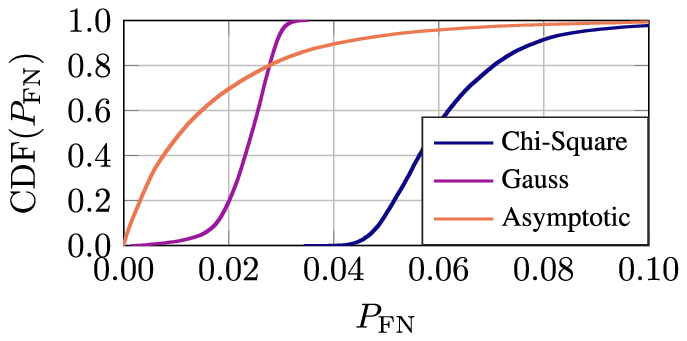}} 
    \subfloat[CDF of the false negatives  $M_{\mathrm{C}}=100$\label{fig:pOverM2-FN100}]{\includegraphics{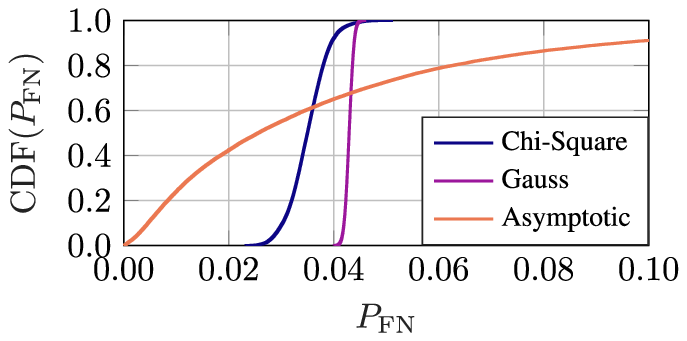}}
    \caption{Empirical expectation of $P_{\mathrm{TN}}$ and $P_{\mathrm{FN}}$ over $M_{\mathrm{C}}$ and CDF of $P_{\mathrm{FN}}$ for two values of $M_{\mathrm{C}}$ with $P_{\mathrm{FN}}^\star=0.05$ with a generalized matrix structure for the three methods provided.}
    \label{fig:pOverM2}
\end{figure*}

Nevertheless, the chi-square approximation is not exact if the channel parameters do not follow an identity-based structure. Hence, we now consider a system with the parameters $\bm{A}_{\mathrm{C}}=a_c\mathrm{diag}\left(\bm{\ddot{a}}\right)$ and $\bm{U}_{\mathrm{C}}=-2a_{\mathrm{C}}\mathrm{diag}\left(\bm{\ddot{u}}\right)$. Thereby, the vectors $\bm{\ddot{a}}$ and $\bm{\ddot{u}}$ are $M_{\mathrm{C}}$-dimensional, and their entries are distributed unitary on $[0.5,1.5]$ and $[0,2]$, respectively. The other parameters remain unchanged from the previous simulation. \figurename~\ref{fig:pOverM2} shows the mean of the probabilities of true and false negatives over $M_{\mathrm{C}}$ together with the corresponding CDF. Therein, \figurename~\ref{fig:pOverM-EPFN2} shows that the chi-square approximation reaches the tolerated $P_{\mathrm{FN}}^\star$ for $M_{\mathrm{C}}=1$, while the achieved number of false negatives changes for larger $M_{\mathrm{C}}$ and can be either slightly too large or slightly too small. Moreover, \figurename{S}~\ref{fig:pOverM2-FN10} and~\ref{fig:pOverM2-FN100} show that the obtained security level has higher variations due to the randomized system parameters and the approximations involved in the chi-square plots. The variations decrease over $M_{\mathrm{C}}$ for both approximations and the Gaussian approximation achieves an almost constant security level for $M_{\mathrm{C}}=100$. 

\section{Conclusion}
In this work, we have designed a Kalman-based physical layer authentication framework, which employs channel and process values jointly to provide authenticity. The security and reliability of the framework have been analyzed based on three approximations, which approximate the cases of few antennas and small measurement vectors, massive SIMO, as well as massive SIMO together with channel hardening. The approximations enable the optimization of the threshold values of the hypothesis test for each received packet individually. While a constant threshold choice provides temporal variations in the security level, the numerical results show that only time-varying threshold values lead to a constant security level. As the thresholds connect the levels of security and reliability, this choice might also lead to temporal variations in the reliability level. The numerical results further validate that in the special cases where no approximations are involved, the proposed scheme can guarantee the desired security level. In the general case, the results are good approximates.

\ifCLASSOPTIONcaptionsoff
  \newpage
\fi

\bibliographystyle{IEEEtran}
\bibliography{IEEEabrv,main}

\end{document}